\documentclass[11pt,twoside,onecolumn]{article}
\usepackage[]{amsmath,amssymb}
\usepackage[]{epsfig}
\newcommand{\be}{\begin{eqnarray}}
\newcommand{\ee}{\end{eqnarray}}

\newcommand\bfk{{\mathbf k}}

\newcommand\bfp{{\mathbf p}}
\newcommand\bfq{{\mathbf q}}

\newcommand\bfx{{\mathbf x}}

\newcommand\sub[1]{_{\rm #1}}

\newcommand\lsim{\lesssim}

\newcommand\half{^{1/2}}

\newcommand{\zetag}{{\zeta\sub g}}
\newcommand\sigmas{{\sigma^2}}

\newcommand{\fnl}{{f\sub{NL}}}

\newcommand\kmax{{k\sub{max}}}
\newcommand\bfkp{{{\bfk}'}}

\newcommand\bfpp{{{\bfp}'}}
\newcommand\tpq{{(2\pi)^3}}
\newcommand\tps{{(2\pi)^6}}

\newcommand\lmone{{L^{-1}}}
\newcommand\zetasig{{\zeta_\sigma}}

\newcommand\vev[1]{{\langle {#1} \rangle}}

\renewcommand\[{\left[}
\renewcommand\]{\right]}

\def\calp{{\cal P}}

\title{\bf Non-Gaussianity from Compositeness}
\author{G.L.~Alberghi\thanks{e-mail: alberghi@bo.infn.it}$\ $
and
R.~Casadio\thanks{e-mail: casadio@bo.infn.it}
\\
\\
{\em Physics Department, Bologna University, and I.N.F.N., Sezione di Bologna}
\\
{\em 
via~Irnerio~46, 40126~Bologna, Italy}}
\begin{document}
\maketitle
\begin{abstract}
By assuming the field seeding the curvature perturbations $ \zeta $ is a dynamically
arising condensate, we are able to derive the relation $ \fnl ^2  \simeq 10^8   H  /   M_c     $
between the non-Gaussianity parameter $ \fnl $ and the ratio of the inflationary scale $ H $
to the cutoff scale $ M_c$ of the effective theory describing the condensate,  thus  relating the 
experimental bound on $ \fnl $ to a bound on  $ M_c$.
\end{abstract}
\pagestyle{plain}
\raggedbottom
\setcounter{page}{1}
\section{Introduction}
\label{intro}
The origin of the structures in the Universe seems to be the primordial curvature
perturbation $\zeta $, present already a few Hubble times before cosmological
scales enter the horizon and come into causal contact.
$\zeta$ is Gaussian within the observational uncertainty and has a a practically
scale-independent spectrum.
Future observation, though, may find a non-Gaussian component.
The usual assumption is that $\zeta $ originates from the vacuum fluctuations
of a light scalar field in an inflationary environment and
these fluctuations are promoted to practically Gaussian classical perturbations
around the time of horizon exit.
By expanding $\zeta $ in powers of the field perturbations, one can see the
linear term is the source of a Gaussian spectrum and the quadratic term can
adequately account for non-Gaussianity~\cite{Lyth-Rodriguez}.
\par
In this article, we will rephrase, within the framework of non-Gussianity,
the results of Ref.~\cite{Alberghi}, where it was assumed that the field at the origin
of  the curvature perturbations is not fundamental but is instead a dynamically arising
condensate.
Explicit examples of the possible realization of the condensation mechanism in the context
of inflation may be found in Refs.~\cite{Giacosa, Alexander}.
In the following we will only assume that such a mechanism is at work.
In this model the correlation functions of a scalar field, usually employed in the
derivation the CMB power spectrum, should therefore be replaced by the correlation
functions for a composite scalar field operator.
The leading behavior of these correlation functions is assumed to reproduce
that of a fundamental scalar field in a de~Sitter space-time, leading to a scale invariant
spectrum for the two-point correlation.
The deviations from a scale-invariant power spectrum, which we interpret as
a sign of the compositeness of the seeding field, can then be consistently related
to a quadratic term in the expansion of the curvature
perturbation $\zeta$ in terms of the seeding field, 
thus providing a link with the literature on non-Gaussianity
\cite{Non-gauss}.
\par
By identifying the corrections obtained in these two different approaches,
one can find a relation between  $ H / M_c $ (the ratio of the  
Hubble scale $ H $ during inflation to the cutoff  $ M_c $ of the effective theory
which describes the scalar condensate) which describes
the amplitude of the correction to the two-point function
and the parameter $ \fnl $ which is usually employed to describe non-Gaussianity.
\par
In the next Section, we briefly review the corrections arising from a fermion
condensate to the spectral index and, in Section~\ref{nonG}, we estimate
the size of possible non-Gaussian contributions.
Finally, we summarize and comment our main results in Section~\ref{conclusions}.
\section{The Condensate Corrections}
\label{condensate}
Let us assume that the condensation mechanism described in
Refs.~\cite{Giacosa, Alexander} is realized in the inflationary expansion
of the Universe, and that the dynamically generated composite scalar field 
is at the origin of the curvature perturbations.
We expect  the composite nature of this field to become manifest in its correlation functions,
and thus in the CMB spectrum.
The equal-time two-point correlation function for a scalar field
$\langle \hat \sigma (\eta, {\bf x}) \hat \sigma (\eta, {\bf x'}) \rangle$ should
be replaced by the two-point function for the condensate $ \bar \psi \psi$,
so that the power spectrum is defined by
\be
\langle (\bar \psi \psi) (\eta, {\bf x}) (\bar \psi \psi)(\eta, {\bf x'})
\rangle 
\equiv \int \frac{dk}{k} 
\frac{\sin k|{\bf x} - {\bf x'}|}{k \, |{\bf x} - {\bf x'}|}
P (k,\eta)
\ .
\label{twopointcond}
\ee
We next assume that the leading behavior of the two-point function is not
modified and produces an almost scale invariant spectrum of perturbations.
For this reason, we factorize the Gaussian leading behavior and take the Fourier transform
of the two-point function $ \tilde \xi (k) $ of the form 
\be
   \tilde \xi (k) \simeq {1 \over k^3} \left(  1 + \delta \tilde \xi \, \right)
\ .
\label{txi}    
\ee
Given the relation
\be
   \calp(k) = {k^3 \over 2 \pi} \,  \tilde \xi (k)  \simeq k^{n_s -1}
\ ,
\ee
one obtains 
\be
    \log\left(  1 + \delta \tilde \xi \right)
    \simeq \left( n_s -1 \right) \log k
    \ ,
\ee
where $n_s$ is the spectral index.
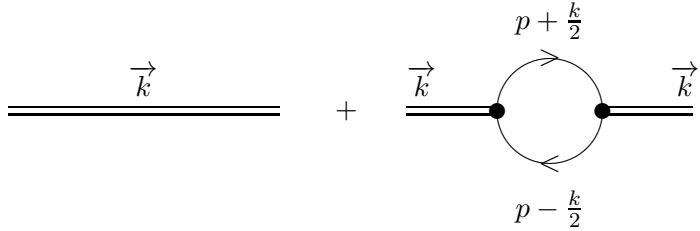
\begin{figure}[t]
\unitlength=1mm
\special{em:linewidth 0.4pt}
\linethickness{0.4pt}
\begin{picture}(91.00,30.00)
\put(20.00,16.50){\line(1,0){36.00}}
\put(20.00,15.50){\line(1,0){36.00}}
\put(38.00,20.00){\makebox(0,0)[cc]{$\overrightarrow k$}}
\put(65.00,16.00){\makebox(0,0)[cc]{$+$}}
\put(92.00,16.00){\circle{14.00}}
\put(99.00,16.00){\circle*{2.00}}
\put(85.00,16.00){\circle*{2.00}}
\put(99.00,16.50){\line(1,0){12.00}}
\put(99.00,15.50){\line(1,0){12.00}}
\put(73.00,16.50){\line(1,0){12.00}}
\put(73.00,15.50){\line(1,0){12.00}}
\put(92.00,28.00){\makebox(0,0)[cc]{$p+{k \over 2}$}}
\put(92.00,3.00){\makebox(0,0)[cc]{$p-{k\over 2}$}}
\put(75.00,20.00){\makebox(0,0)[cc]{$\overrightarrow k$}}
\put(110.00,20.00){\makebox(0,0)[cc]{$\overrightarrow k$}}
\put(92.00,23.00){\makebox(0,0)[cc]{$>$}}
\put(92.00,9.00){\makebox(0,0)[cc]{$<$}}
\end{picture}
\caption{Diagrammatic representation of Eq.~\eqref{loops}.
Double lines correspond to composite scalars of momentum $ k $.}
\label{diagram}
\end{figure}
\par
We now need to determine the form of the modification of the Fourier
transform of the two-point function $ \delta \tilde \xi (k) $ which corresponds
to the four-point function for the fermion field $ \psi $ in the particular case where
two fermions form a composite of ingoing momentum 
$ k $ and two form a composite of outgoing momentum $ k $. 
This way the four-point function for the fermion fields reduces to the two-point function
for the composite scalars.
If we look at the flat space case (see Ref.~\cite{Compositeness1}), we see that,
in the momentum representation,  the inverse propagator for a composite scalar particle
is modified as (see Fig.~\ref{diagram})
\be
 \Gamma (k) =\delta Z \left( \Gamma_0 (k) +  \delta \, \Gamma (k) \right)  \equiv \delta Z  (m^2 + k^2)  \, I (k)
 \ , 
\ee
where $ \Gamma_0 (k)  $ is the free scalar propagator, $ \delta Z $ the wave-function renormalization,
$ I(k) $ the form factor, and 
\be      
     \delta \, \Gamma (k)   \simeq  \int  {d ^4 p \over (2 \pi)^4 } \,  {\rm tr} \left[   
     (  \kern+0.1em  /\kern-0.55em p + \kern+0.1em  /\kern-0.55em k/2 + im )^{-1} 
      ( \kern+0.1em  /\kern-0.55em p -  \kern+0.1em  /\kern-0.55em k/2 + im )^{-1}  \right ]
      \ . 
\label{loops}
\ee 
The form factor $ I(k) $ is related to the non-local component of the propagator which is 
given by the fermion loop.
We also note that a wave function renormalization for the composite field is required in order to 
get a nonsingular behavior of the two point function.
On the other hand, the assumption that the leading order of the two-point function for the composite field
reproduces the two-point function of a scalar, is equivalent to assuming that the only
nonlocal term describing the modification of the two-point function comes from the fermion loop.
By looking at Ref.~\cite{Compositeness2}, one can see that loop diagrams
were evaluated in the cosmological context in order to describe the influence of higher order 
quantum effects on the CMB spectrum.
A conclusion one can draw from Refs.~\cite{Compositeness2} is that a logarithmic correction
to the Fourier transform of the two-point function is  expected from the fermion loop.
As in any effective theory, details which depend on the high-energy physics enter only through
non-renormalizable operators, which will here be suppressed by the compositeness scale.
Therefore, in the absence of a special argument to the contrary, we must expect that the
leading effects from compositeness are proportional to  $ A\equiv H / M_c $ where $M_c$ is the 
cutoff of the effective theory above which the compositeness effects become unobservable.
If the condensation scale is the Planck mass $M_{\rm P} $, as implied by the models described in
Refs.~\cite{Giacosa,Alexander}, then $A \sim 10^ {-5} $, with lower values of $ M_c $ implying
bigger values for $A$. 
\par
\begin{figure}
\label{grafico1}
\centerline{
\epsfxsize=8cm \epsfbox{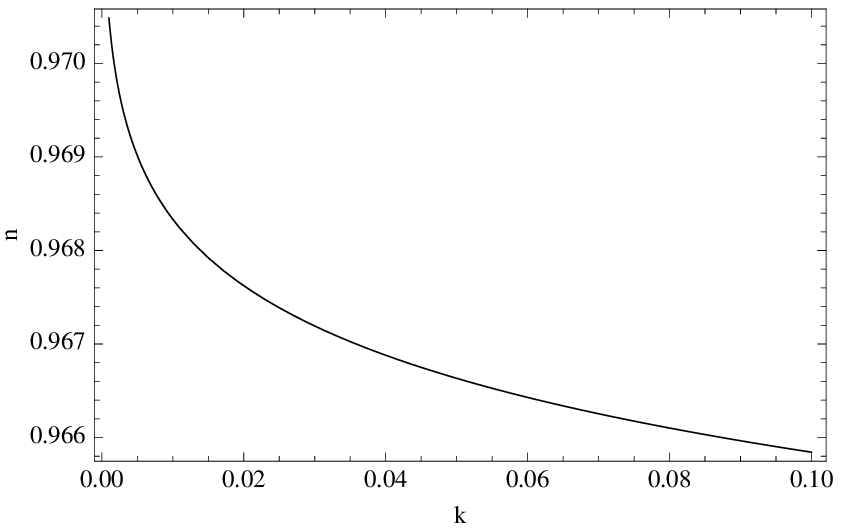}
\
$\quad$
\epsfxsize=8cm \epsfbox{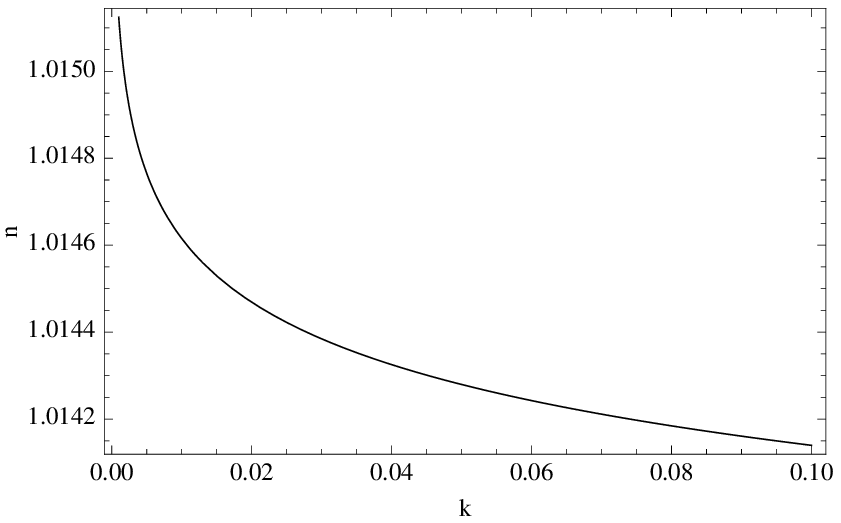}
}
\caption{Spectral index $n_s$ for $ A = 10^{-2}$ and $ B = 10^{-6} \,  \rm{Mpc}^{-1}$ 
(left panel)
and for $ A = -10^{-2}$ and $ B = 10^{-6} \, \rm{Mpc} ^{-1} $ (right panel).}
\end{figure}

To summarize, our ansatz for the correction to $ \tilde \xi $ in Eq.~\eqref{txi}
is of the form
\be
\label{modification}
      \delta \tilde \xi =    A\, \log \left(   {B \over k} \right)
      \ ,
\ee 
where $ B $ is, in general, the unit $ k $ is measured (or renormalization scale) and
which we will relate to the Hubble horizon in the following Section. 
Note also that we here allow the coupling constant for the dynamically generated condensate 
to be more general than that described in Refs.~\cite{Giacosa, Alexander}, where the factor
$A $ would naturally be of the order $ H / M_{\rm P} $.
Eq.~\eqref{modification} leads to a spectral index of the form
\be
   n_s  \simeq  1 + {  \log  \left[ 1 + A \log  \left(  {B \over k }\right) \right] \over \log k }
   \ ,
\ee
whose behavior is shown in Fig.~\ref{grafico1} for positive and negative $A$.
In the first case the spectral index is smaller than one for all $ k $ and a red spectrum is generated,
whereas in the second case it is larger than one and a blue spectrum results. 
\section{Non-Gaussianity}
\label{nonG}
We now consider the previous results in the framework of the 
formalism used to extract the non-Gaussian features of the CMB spectrum.
Following Ref.~\cite{Lyth2005}, we assume the curvature perturbation $ \zeta $ is the sum of two terms
\be
\zeta(\bfx) &=&  \zetag(\bfx) + \zetasig(\bfx)
\ee
where $\zetag$ and $  \zetasig $ are respectively its Gaussian and non-Gaussian component.
One can consistently express the non-Gaussian part as the sum of a linear and a quadratic term in the field $ \sigma $ as
\be 
 \zeta_{\sigma}(\bfx)  = \sigma(\bfx)  -\alpha\,  \fnl \left( \sigma^2(\bfx) -
\vev{\sigma^2(\bfx)} \right)
\ ,
\label{corrchis} 
\ee
where $\sigma(\bfx) $ now represents the condensate field and $ \alpha $ is a coefficient
of order one (which depends on the convention adopted).
There are in principle two possibilities: 
({\em i\/})~the field $ \sigma $ which describes the scalar condensate
is the dominating field in the generation of the curvature perturbations.
In this case the linear part in (\ref{corrchis}) accounts for the Gaussian part $\zeta_g $ of the spectrum
and the quadratic term accounts for the non-Gaussian part;
({\em ii\/})~the field $ \sigma $ is sub-dominant in the generation of the curvature
perturbation $\zeta$ (in this case $ \zeta_g $ is generated by another field),
so that we can neglect the linear part in (\ref{corrchis}) and concentrate on the quadratic term.
In any case we  have
\be
&&P_\zeta (k)=P_\zetag(k) +  P_\sigmas(k)
\ .
\label{phis1}
\ee
We first evaluate the non-Gaussian part of the spectrum coming from
the quadratic term in Eq.~\eqref{corrchis}.
Fourier components of $\sigmas$ are given by
\be
(\sigmas)_\bfk = \frac1\tpq \int d^3q \, \sigma_\bfq \sigma_{\bfk-\bfq}
\label{phisk}
\,.
\ee
For non vanishing $\bfk$ and $\bfk'$
\be
\vev{(\sigmas)_\bfk (\sigmas)_\bfkp} &=& \frac1\tps \int d^3p \,d^3p'\;
\vev{\sigma_\bfp \sigma_{\bfk-\bfp} \sigma_\bfpp \sigma_{\bfkp-\bfpp} }
\nonumber\\
&=& 2\delta^{(3)}(\bfk+\bfk')\int d^3p \, P_\sigma(p) P_\sigma(|\bfk-\bfp|)
\ ,
\nonumber
\ee
and, taking $\calp_\sigma$ scale-independent~\cite{myaxion}, 
\be
\calp_\sigmas(k) = \frac{k^3}{2\pi} \calp_\sigma^2 \int_\lmone 
\frac{d^3p}{ p^3 |\bfp-\bfk|^3 }
\label{psigs}
\,.
\ee
The subscript $\lmone$ indicates that the integrand 
is set equal to zero in a sphere of radius 
$\lmone$ around each singularity,
and   the discussion makes sense only for $\lmone \ll k \ll \kmax$.
In this regime one finds~\cite{myaxion} 
\be
\calp_\sigmas(k) = 4
\calp_\sigma^2 \ln(kL) \label{exact}
\ .
\ee
If we plug this result back into Eq.~\eqref{phis1} we obtain
\be
  P_\zeta (k)=P_\zetag(k) +  P_\sigmas(k) =  
   P_\zetag(k) \left[  1 + 8 \, \alpha^2\,   \fnl ^2  \, \calp_\sigma\, \log (kL) 
   \right]
   \ .
   \label{spectrum-ng}
\ee
We can now compare this result with our discussion of compositeness effects in the
spectrum of perturbations.
The first thing to notice by looking at Eqs.~\eqref{modification} and~\eqref{spectrum-ng}
is that they coincide in producing a logarithmic correction to the Gaussian part.
\par
We note that  the correction arising from Eq.~\eqref{corrchis}
has a definite sign, since the factor in front of the logarithmic correction is given by
$ 8 \,\calp_\sigma \alpha ^2\, \fnl^2>0 $, and thus maps into the case $ A < 0 $ of
Eq.~\eqref{modification}, which gives a blue correction to the spectrum.
However the size of this correction is very small, of the order of one part on $ 10^{-5}$
and its effect will not appear in the experimental measurements of the spectral index.
Nonetheless we will see that its effects can be detectable in the measurements of the bispectrum.
Let us now consider the two possibilities described earlier.
The first one is that the linear term in the expansion (\ref{corrchis}) is responsible for the 
Gaussian part of the spectrum. As we have seen the correction coming from the non-Gaussian term
provides a negligible correction and we have to assume that a potential is generated for the condensate
in order to account for the deviation from $ n_s = 1 $. If this is the case 
we can take $ \calp_\sigma \simeq \calp_{\zeta_g} \simeq 10^{-9} $.
By identifying  the amplitude of the logarithmic correction
for the composite scalar case with the contribution due to the quadratic term
we are able to make an estimate of the magnitude of the bispectrum $B_g$ defined by
\be
\vev{ g_{\bfk_1}\, g_{\bfk_2}\, g_{\bfk_3} }
=
\tpq\, \delta^{(3)}(\bfk_1+\bfk_2+\bfk_3)\,B_g(k_1,k_2,k_3) 
\ .
\ee
In fact if the curvature perturbation has  the form~\eqref{corrchis}, its bispectrum is given to
leading order by~\cite{spergel}
\be
 B_\zeta(k_1,k_2,k_3)
=
-2\,\alpha\,
\fnl \[ P_\zeta(k_1)\, P_\zeta(k_2) +\,{\rm cyclic} \] 
\label{fnldef}
\ .
\ee
(Only the  term linear in $\fnl$ is kept, which is justified because the second term of Eq.~\eqref{corrchis}
is much smaller than the first term.) 
Current observations~\cite{komatsu,new} give  $|\fnl|\lesssim 100$,
which makes the non-Gaussian fraction of $\zeta$ less than $100 \, \calp_\zeta\half\sim 10^{-3}$.
Absent a detection, PLANCK~\cite{planck} will bring this down to roughly $|\fnl|\lsim 1$.
Now, identifying the correction to the power spectrum due to the non-Gaussian term
with that due to compositeness yields
 \be
A= 8\, \alpha ^2\, \fnl ^2 \,  \calp_ {\zeta_g}   
\ ,
\label{Afnl}
\ee
which means that 
\be
     \fnl ^2  \simeq 10^8   {H  \over  M_c }   
\ .   
\ee
This is our prediction for the value of $ \fnl $, and by taking  $ A \lesssim 10^{-4} $,
we obtain $ \fnl \lesssim 100 $ which is the current experimental bound.  
Conversely, the bound $ |\fnl| \lesssim 100 $ may be interpreted as a constraint on $ A $
in the form $ A \lesssim 10^{-4}\, \alpha ^2  \simeq  10^{-4}$.
This yields a lower bound on the effective
scale $M_c\gtrsim 10^{-1}\, M_P  $.
\par
The second possibility is that the non-Gaussian contribution is sub-dominant.
In this case instead of equating $ \calp_{\zeta} $ with $ \calp _\sigma $ in eq. (\ref{spectrum-ng})
one can express $ \calp _ \sigma = r \calp_\zeta $ which describes the ratio between the
non-Gaussian part and the Gaussian part of the spectrum, and is related to the so called
non-Gaussian fraction $ r _ {\zeta_\sigma} \equiv ( \calp_{\sigma^2} / \calp_\zeta)^{1 /2} $.
In this case the relation (\ref{Afnl}) becomes  $ A= 8\, r \, \alpha ^2\, \fnl ^2 \,  \calp_ {\zeta_g}   $,
and for $ r = 10^{-1} $, one has that
$ \fnl \simeq 100 $ implies $ A = H / M_p$ as expected in the case where the mechanism of
cosmological condensation described in \cite{Giacosa, Alexander} is at work. 

\section{Conclusions}
\label{conclusions}
By assuming the field seeding the non-Gaussian component of curvature perturbations $ \zeta $ is a dynamically
generated condensate, we derived the relation~\eqref{Afnl} for the parameter $ \fnl $ in terms of the ratio of the inflationary scale $ H $ to the cutoff
scale $ M_c$ of the effective theory describing the condensate. The condensate corrections affect the bispectrum according to
Eq.~\eqref{fnldef}, thus inducing the bound $M_c\gtrsim 10^{-1} \, M_P$. If the condensate field is sub-dominant it is shown that the 
mechanism of cosmological condensation of  Refs. \cite{Giacosa, Alexander} can produce $ \fnl \simeq 100 $.

\label{conc}
\par
\end{document}